\documentclass[prb,twocolumn,superscriptaddress,showpacs]{revtex4}
\usepackage{amsmath,amssymb,graphicx}
\usepackage{color}

\begin{document}

\title{Vertical cavity surface emitting terahertz laser}
\author{A.V. Kavokin}
\affiliation{Spin Optics Laboratory, St-Petersburg State University, 1, Ulianovskaya,
198504, Russia and\ School of Physics and Astronomy, University of
Southampton, Highfield, Southampton SO17 1BJ, United Kingdom}
\author{I.A. Shelykh}
\affiliation{Science Institute, University of Iceland, Dunhagi-3, IS-107, Reykjavik,
Iceland and Division of Physics and Applied Physics, Nanyang Technological
University 637371, Singapore}
\author{T. Taylor}
\affiliation{School of Physics and Astronomy, University of Southampton, Highfield,
Southampton SO17 1BJ, United Kingdom}
\author{M.M. Glazov}
\affiliation{Ioffe Physical-Technical Institute of RAS, 194021 St.\ Petersburg, Russia}
\date{\today}

\begin{abstract}
Vertical cavity surface emitting terahertz lasers can be realized in
conventional semiconductor microcavities with embedded quantum wells in the
strong coupling regime. The cavity is to be pumped optically at half the
frequency of the $2p$ exciton state. Once a threshold population of $2p$
excitons is achieved, a stimulated terahertz transition populates the lower
exciton-polariton branch, and the cavity starts emitting laser light both in
the optical and terahertz ranges. The lasing threshold is sensitive to the
statistics of photons of the pumping light.
\end{abstract}

\maketitle

Creation of efficient sources of terahertz (THz) radiation is of high
importance for various fields of modern technology, including information
transfer, biosensing, security and others\cite{Dragoman}. These applications
are currently limited due to the lack of compact and reliable solid state
sources of THz radiation. The main obstacle preventing the creation of such
a source is the low rate of spontaneous emission of THz photons: according
to Fermi's golden rule this rate is proportional to the cube of the
frequency and for THz transitions should be roughly tens of inverse
milliseconds, while lifetimes of crystal excitations typically lie in the
picosecond range\cite{Duc,Doan}. The attempted strategies to improve this
ratio include using the Purcell effect\cite{Gerard,Todorov} by embedding the
sample inside a THz cavity, or using the cascade effect in quantum cascade
lasers\cite{Faist} (QCLs). Nevertheless, until now QCLs in the spectral
region around 1 THz remain costly, have macroscopic dimensions, are
short-lived and still show a quantum efficiency of less than 1\%.

Recently, it has been proposed that the emission rate of THz photons may be
increased by bosonic stimulation if the THz transition feeds a condensate of
exciton-polaritons. In Refs.\onlinecite{Kavokin1,SavenkoPRL,delValle} the
authors suggest using the transition between the upper and lower polariton
branches in a semiconductor microcavity in the polariton lasing regime. The
radiative trasition between two polariton modes, accompanied by emission of
a THz photon, may become allowed if an electric field mixing polariton and
dark exciton states is applied to the cavity. THz photons would be emitted
in the plane of the cavity, and a supplementary lateral THz cavity would be
needed to provide the positive feedback.

Here we propose an alternative model of a microcavity based THz laser, which
has significant advantages compared to one based on the transition between
two polariton branches. We propose using two-photon pumping of a $2p$
exciton state, as has been realised already in GaAs based quantum well
structures\cite{Cingolani, Schemla}. The direct transition to or from the $2p
$ exciton state with emission or absorption of a single photon is forbidden
by optical selection rules. Instead, a $2p$ exciton can radiatively decay to
the lower exciton-polariton mode formed by the $1s$ exciton and cavity
photon. This transition is accompanied by emission of a THz photon\cite%
{Garro}. The inverse process (THz absorption by a lower polariton mode with
excitation of a $2p$ exciton) has been recently observed experimentally \cite%
{KochCondmatt}. The THz transition from the $2p$ state pumps the lowest
energy exciton-polariton state, which eventually leads to the polariton
lasing effect, widely discussed in the literature \cite{Christopoulos}. A
macroscopic occupation of the lowest energy polariton state stimulates
emission of THz photons, so that, in the polariton lasing regime, the cavity
would emit one THz photon for each optical photon emitted by the polariton
laser, ideally.

\begin{figure}[tbp]
\includegraphics[width=9cm]{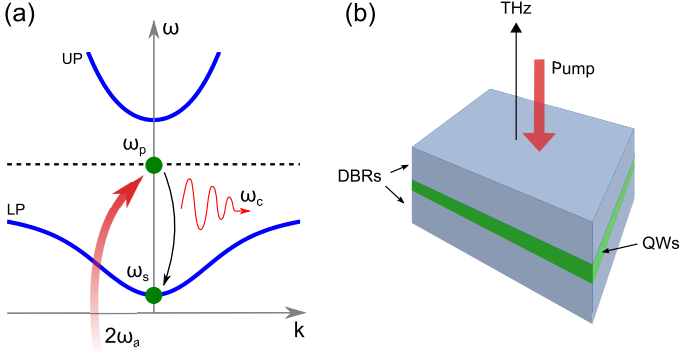}
\caption{(a) Schematic of the polariton dispersion relation, showing the
lower-polariton (LP) and upper-polariton (UP) branches, as well as the $2p$
exciton state with frequency $\protect\omega_p$. The pump frequency, $%
\protect\omega_a$, is half that of the $2p$ exciton state, as discussed in
the text. (b) The structure considered in this work: a semiconductor
microcavity consisting of an active layer containing quantum wells (QWs)
sandwiched between two distributed Bragg reflectors (DBRs). The structure is
pumped vertically, i.e.\ in the direction perpendicular to the microcavity
plane, and the resulting THz emission from the cavity is in the same
direction.}
\label{schematic}
\end{figure}

The design of the laser and the involved energy levels are illustrated
schematically in Figure 1. This design has two crucial advantages with
respect to that previously considered \cite{Kavokin1}: it allows for
operation with an optically allowed THz transition, and it provides vertical
emission of THz photons. The whole structure is microscopic; no waveguides
or THz cavities are needed. This latter point is also a significant
advantage with respect to the quantum cascade laser, which operates in the
wave-guide geometry. Moreover, the suggested scheme is very interesting from
a fundamental point of view, since, as shown below, the threshold of the
proposed laser appears to be sensitive to the statistics of photons of the
pumping light. Here we present the quantum model of this laser based on the
Liouville equation for the density matrix describing the $2p$ exciton state,
the lower polariton mode, and optical and THz photons.

The system under study consists of 2$p$ excitons and 1$s$ exciton-polaritons
interacting with an external electromagnetic field and phonons in a
semiconductor microcavity. Hence, our consideration involves both coherent
and incoherent processes. Consequently, it is convenient to split the total
Hamiltonian of the system into two parts, 
\begin{equation}
H=H_{c}+H_{d},
\end{equation}%
where the part $H_{c}$ includes all types of coherent processes in our
system and part $H_{d}$ describes the decoherence in the system due to
exciton interactions with acoustic phonons, modelled as a classical
reservoir \cite{Magnusson,Savenko}.

The equation for the density matrix can be written in the following form: 
\begin{equation*}
\frac{d\varrho }{dt}=\frac{i}{\hbar }\left[ \varrho ,H_{c}\right] +\hat{L}%
\varrho,
\end{equation*}%
where $\hat{L}$ is a Lindblad superoperator defined below. The incoherent
part of the Hamiltonian can be divided into two parts, $H_{d}=%
\sum_i(H^{+}_i+H^{-}_i)$, where $H^{+}_i$ creates an excitation in the
quantum system (and thus annihilates an excitation in the classical
reservoir), and $H^{-}_i$, conversely, annihilates an excitation in the
quantum system and creates an excitation in the classical reservoir. The
subscript index $i$ numerates different types of decoherence processes
in the system.

The Lindblad terms can be then written as \cite{Savenko,Ostatnicky} 
\begin{eqnarray}
\hat{L}\varrho  &=&\frac{\delta (\Delta E)}{\hbar }\sum_{i}\left\{ \left(
H_{i}^{+}\rho H_{i}^{-}+H_{i}^{-}\varrho H_{i}^{+}\right) -\right.  \\
&&-\left. \left( H_{i}^{+}H_{i}^{-}+H_{i}^{-}H_{i}^{+}\right) \varrho
-\varrho \left( H_{i}^{+}H_{i}^{-}+H_{i}^{-}H_{i}^{+}\right) \right\} , 
\notag
\end{eqnarray}%
where $\delta (\Delta E)$ accounts for energy conservation, and in realistic
calculations should be taken as an average inverse broadening of the states
in our system, $\delta (\Delta E)\rightarrow \zeta ^{-1}$. In the particular
system which we consider, the coherent and incoherent parts of the
Hamiltonian can be written as 
\begin{eqnarray}
H_{c} &=&\epsilon _{p}p^{+}p+\epsilon _{s}s^{+}s+\epsilon
_{a}a^{+}a+\epsilon _{T}c^{+}c, \\
H_{d} &=&\sum_{i=1,2}(H_{i}^{+}+H_{i}^{-}),
\end{eqnarray}%
where 
\begin{eqnarray}
H_{1}^{+} &=&gp^{+}a^{2}, \\
H_{1}^{-} &=&gpa^{+2}, \\
H_{2}^{+} &=&Gp^{+}sc, \\
H_{2}^{-} &=&Gps^{+}c^{+}.
\end{eqnarray}%
Here $p$ and $s$ denote annihilation operators of the $2p$ exciton and the
lowest energy 1$s$ polariton states, respectively, $a$ is the annihilation
operator for laser photons exciting the $2p$ exciton state, and $c$ is an
annihilation operator for THz photons produced by the $2p\rightarrow 1s$
transition. $\epsilon _{p}$, $\epsilon _{s}$, $\epsilon _{a}$ and $\epsilon
_{T}$ are the energies of the $2p$ exciton, $1s$ exciton-polariton, pump
photon ($2\epsilon _{a}=\epsilon _{p}$) and THz photon, $\epsilon
_{T}=\epsilon _{p}-\epsilon _{s}$, respectively. Constants $g$ and $G$
define the strengths of the two-photon absorption and $1s\rightarrow 2p$
radiative transitions, respectively. While the $1s\rightarrow 2p$ transition
strength is determined by the dipole matrix elements, the main contribution
to the two-photon $2p$ excitation comes from processes with intermediate
states in the valence or conduction bands~\cite{ivchenkopikus}. We assume
that the upper polariton state is higher in energy than the 2$p$ exciton
state, and thus can be excluded from consideration. The classical reservoir
consists of the photons of the external laser light and THz photons. Note,
that the coherent part of our Hamiltonian describes only the free modes of
the system and thus is irrelevant to the kinetic equations.

Now, for the occupancy of the 2$p$ exciton mode one can obtain after
straightforward algebra: 
\begin{eqnarray}
\frac{dN_{p}}{dt} &=&Tr\left\{ p^{+}p\frac{d\varrho }{dt}\right\} =\frac{2}{%
\hbar \zeta }\text{Re}\{\text{Tr}\left( \varrho \lbrack
H^{-};[p^{+}p;H^{+}]]\right) \}=  \notag \\
&=&W_{g}\left[ \frac{1}{2}\langle a^{+2}a^{2}\rangle -\langle
p^{+}p(2a^{+}a+1)\rangle \right] +  \notag \\
&&+W_{G}\left[ \langle s^{+}sc^{+}c\rangle -\langle
p^{+}p(s^{+}s+c^{+}c+1)\rangle \right] ,  \label{dNpdt}
\end{eqnarray}%
where $\langle ...\rangle =Tr\left\{ ...\varrho \right\} $ denotes averaging
with the appropriate density matrix $\varrho $, $W_{g}=4g^{2}/\hbar \zeta$
and $W_{G}=2G^{2}/\hbar \zeta $. The equations for the polariton mode
occupancy $N_{s}=\langle s^{+}s\rangle $, and the terahertz mode occupancy $%
N_{c}=\langle c^{+}c\rangle $ are analogous to those for $N_{p}$. The
occupancy of the pumping mode $N_{a}=\langle a^{+}a\rangle $ is defined by
the intensity of the external pump and we do not need to write an independent
dynamic equation for it. The same holds true for the higher order
correlators involving pump operators, e.g. $\langle a^{+2}a^{2}\rangle $. It
follows from Eq.~(\ref{dNpdt}) that the dynamic equations for the
occupancies of the modes contain quantum correlators of fourth order, such
as $\langle s^{+}sc^{+}c\rangle $. For them one can also write the dynamic
equations analogous to Eq.~(\ref{dNpdt}), which would contain correlators of
sixth order. Proceeding further, one would obtain an infinite chain of
coupled equations for the hierarchy of correlators.

In order to solve this chain of equations one needs to truncate the
correlators at some stage. Here we use the mean-field approximation, which
consists in truncation of the fourth-order correlators into products of
second-order ones. One can approximate $\langle p^{+}pa^{+}a\rangle \approx
\langle p^{+}p\rangle \langle a^{+}a\rangle =N_{p}N_{a},\langle
s^{+}sc^{+}c\rangle \approx N_{s}N_{c},...$ etc.

A particular analysis is needed for the truncation of the correlator $%
\langle a^{+2}a^{2}\rangle $ containing four operators corresponding to the
pumping mode. Using the definition of the second order coherence $g^{(2)}(0)$%
, it can be represented as 
\begin{equation}
\langle a^{+2}a^{2}\rangle =g^{(2)}(0)N_{a}^{2}.  \label{g2}
\end{equation}%
Finally, the closed set of equations of motion describing the dynamics of
our system reads 
\begin{eqnarray}
\frac{dN_{p}}{dt} &=&-\frac{N_{p}}{\tau _{p}}+W_{g}\left[ \frac{g^{(2)}(0)}{2%
}N_{a}^{2}-N_{p}(2N_{a}+1)\right] +  \label{dNpdt1} \\
&&+W_{G}\left\{ N_{s}N_{c}\left( N_{p}+1\right) -N_{p}\left( N_{s}+1\right)
\left( N_{c}+1\right) \right\} ,  \notag \\
\frac{dN_{s}}{dt} &=&-\frac{N_{s}}{\tau _{s}}- \\
&&-W_{G}\left\{ N_{s}N_{c}\left( N_{p}+1\right) -N_{p}\left( N_{s}+1\right)
\left( N_{c}+1\right) \right\} ,  \notag \\
\frac{dN_{c}}{dt} &=&-\frac{N_{c}}{\tau _{c}}- \\
&&-W_{G}\left\{ N_{s}N_{c}\left( N_{p}+1\right) -N_{p}\left( N_{s}+1\right)
\left( N_{c}+1\right) \right\} ,  \notag  \label{dNcdt1}
\end{eqnarray}%
where we have introduced the non-radiative lifetime of the 2p exciton state $%
,\tau _{p}$, and lifetimes of lower polaritons and THz photons, $\tau _{s}$
and $\tau _{c}$, respectively.

Let us analyze in more detail the kinetic equation for the $2p$ state
occupation $N_{p}$. The terms describing incoming and outcoming scattering
rates for this state can be written as 
\begin{equation*}
(1+N_{p})\langle {a^{+}}^{2}a^{2}\rangle -N_{p}\langle a^{2}{a^{+}}%
^{2}\rangle .
\end{equation*}%
As we already mentioned in Eq.~(\ref{g2}), the incoming rate is proportional
to $g^{(2)}(0)N_{a}^{2}$. The outgoing rate can be recast as $\propto
\langle a^{2}(a^{+})^{2}\rangle =g^{(2)}(0)N_{a}^{2}+4N_{a}+2$. Hence, all
terms containing $N_{a}^{2}N_{p}$ cancel each other and Eq. (\ref{dNpdt1})
holds. Note, that for processes of non-degenerate two-photon absorption or
emission neglected here, two-photon emission processes where photon
frequencies are different can be included in $\tau _{p}$, the decay rate of
the $2p$ state. The scattering rates have standard form $%
(1+N_{p})N_{a_{1}}N_{a_{2}}-N_{p}(N_{a_{1}}+1)(N_{a_{2}}+1)$, independent of
the photon field statistics~\cite{enaki}. In what follows we assume that $%
N_{a}\gg 1$ making it possible to neglect all other modes.

\begin{figure}[tbp]
\includegraphics[width=8.5cm]{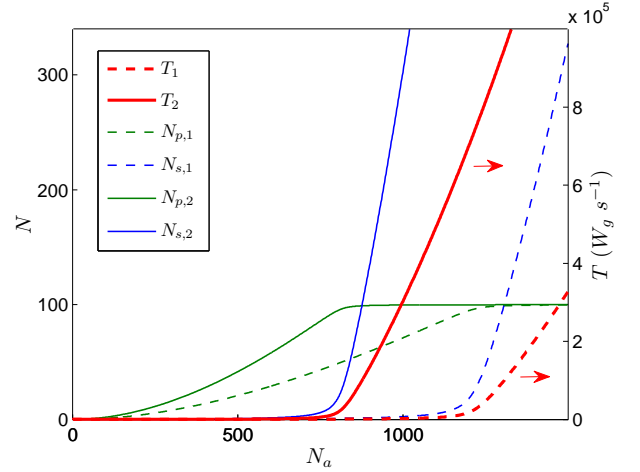}
\caption{Occupation of the $p$- and $s$-states in the steady state (left
axis) and terahertz generation rate (right axis), as a function of pumping
mode occupancy. The subscript 1 or 2 refers to the value of $g^{(2)}(0)$
taken. $W_g = 1, W_G = 10, 1/\protect\tau_s = 1000, 1/\protect\tau_p = 5000 $%
, in units of $W_g$.}
\label{thresholds}
\end{figure}

The key feature of Eq.~(\ref{dNpdt1}) is that the pumping term $%
W_{g}g^{(2)}(0)N_{a}^{2}$ contains the second order coherence of the pump,
i.e.\ strongly depends on its statistics. For a laser pump corresponding to
the coherent state this statistics is Poissonian and $g^{(2)}(0)=1$. On the
other hand different radiation sources may provide different statistics of
pump photons and different values of $g^{(2)}(0)$. For example, for a
thermal pump $g^{(2)}(0)=2$. Different values of $g^{(2)}(0)$ lead to the
different lasing thresholds in the system. Note that dependence of
two-photon optical processes on the second order coherence of light has been
discussed by Ivchenko in Ref.~\onlinecite{Ivchenko}. In Figure 2, we plot the
steady state solutions for $N_{p}$ and $N_{s}$, assuming that the terahertz
mode occupation $N_{c}$ is zero, for the cases of coherent and thermal
pumps. For a qualitative understanding, we note that the threshold is
reached if $N_{s}\sim 1$, or equivalently when $N_{p}\sim 1/\tau _{s}W_{G}$,
since this signals the onset of Bose stimulation of the transition from the $%
2p$-state to the lower polariton, accompanied by terahertz emission. The
terahertz generation rate is given by $T=W_{G}N_{p}(N_{s}+1)\approx 1/\tau
_{s}N_{s}$, and therefore shows the same threshold behaviour with pumping
intensity as $N_{s}$, as can also be seen in Figure 2. We observe that the
threshold is higher in the case of a coherent pump, which may be understood
in terms of the enhanced losses from the $2p$-state due to stimulated
two-photon emission.

The dynamics of switching the system to polariton and terahertz lasing is
essentially governed by the dynamics of the correlator $\langle
a^{+2}a^{2}\rangle $, i.e. by the product of squared intensity and second
order coherence of the pumping light. The efficiency of two-photon
absorption $W_{g}$ and the $1s$ polariton lifetime $\tau _{s}$, dependent on
the quality factor of the optical cavity, are two main factors which govern
the threshold to lasing. The polarisation of pumping defines the
polarisation of emission, and the frequency of THz radiation may be tuned by
changing the exciton-photon detuning in the microcavity. Interestingly, THz
lasing in our system is possible in the absence of a THz cavity. In order to
demonstrate this, we have taken the terahertz photon lifetime $\tau _{c}=0$ in
the calculation shown in Figure 2. One can see that THz lasing begins
simultaneously with polariton lasing.  The stimulated emission of THz
photons is achieved due to the bosonic stimulation of this transition by
occupation of the final ($1s$ polariton) state. The system emits THz
radiation through the Bragg mirrors of the microcavity, which are transparent
at THz frequencies. 

In conclusion, we have developed a quantum theory of vertical cavity surface
emitting terahertz lasers. The vertical terahertz lasing coexists with
polariton lasing and, consequently, is characterised by a low pumping
threshold. Interestingly, the value of the threshold is strongly dependent
on the statistics of photons of the pumping light. For practical realization
of compact terahertz lasers operating at room temperature, microcavities
based on wide band gap semiconductors (GaN or ZnO) would seem to be the most
advantageous. The resonant two photon pumping of 2$p$ exciton states in such
a cavity may be assured by a conventional vertical cavity light emitting
diode (VCLED) emitting in red. This prospective structure would consist of a
GaN or ZnO microcavity grown on the top of a GaAs based VCLED structure. As
emission of coherent terahertz light coexists with the polariton lasing in
the optical frequency range in this scheme, one can use the visible blue or
green light produced by the polariton laser as a marker for the terahertz
light beam, which may be important for applications in medicine, security
control, etc.

AVK thanks Alberto Bramati, Elisabeth Giacobino, Elena Del Valle and Fabrice
Laussy for fruitful discussions. This work has been supported by the
visitors program of the International Institute of Physics (Natal, Brazil),
and EU IRSES project "POLAPHEN". IAS acknowledges the support from Rannis
``Center of Excellence in Polaritonics''. MMG was partically  supported by
RFBR.


\begin{thebibliography}{99}
\bibitem{Dragoman} D. Dragoman, M. Dragoman, \emph{Progr. in Quant.
Electronics} \textbf{28}, 1 (2004).

\bibitem{Duc} H. T. Duc et al, Phys. Rev. B \textbf{74}, 165328 (2006).

\bibitem{Doan} T. D. Doan et al, Phys. Rev. B \textbf{72}, 085301 (2005).

\bibitem{Gerard} J.-M. Gerard and B. Gayral, \emph{Journ. Lightwave Technol.}
\textbf{17}, 2089 (1999).

\bibitem{Todorov} Y. Todorov \emph{et al}, \emph{Phys. Rev. Lett.} \textbf{99%
}, 223603 (2007).

\bibitem{Faist} R.F. Kazarinov, and R.A. Suris, \emph{Sov. Phys.
Semiconductors} \textbf{5}, 707 (1971); J. Faist \emph{et al}, \emph{Science}%
, \textbf{264,} 553 (1994); E. Normand \emph{et al}, \emph{Laser Focus World}%
, \textbf{43,} 90 (2007).

\bibitem{Kavokin1} K. V. Kavokin \textit{et al}, Appl. Phys. Lett. \textbf{97%
}, 201111 (2010).

\bibitem{SavenkoPRL} I.G. Savenko, I.A. Shelykh and M.A. Kaliteevski, Phys.
Rev. Lett. \textbf{107}, 027401 (2011).

\bibitem{delValle} E. del Valle and A. V. Kavokin, Phys. Rev. B \textbf{83},
193303 (2011).

\bibitem{Cingolani} I.M. Catalano \textit{et al}, Phys. Rev B \textbf{40},
1312 (1989).

\bibitem{Schemla} R. A. Kaindl, D. Hagele, M. A. Carnahan, and D. S. Chemla,
Phys. Rev. B \textbf{79}, 045320 (2009).

\bibitem{Garro} N. Garro, S.P. Kennedy, A.P. Heberle, R.T. Phillips, Physica
Status Solidi (b), \textbf{221}, 385 (2000).

\bibitem{KochCondmatt} J.L. Tomaino \textit{et al}, arXiv:1112.2185 (2011).

\bibitem{Christopoulos} see e.g. S. Christopolous \textit{et al}, Phys. Rev.
Lett.\textbf{\ 98}, 126405 (2007); A. Das \textit{et al, }Phys. Rev. Lett.%
\textit{\ }\textbf{107}, 066405 (2011).

\bibitem{Magnusson} E. B. Magnusson \textit{et al}, Phys. Rev. B \textbf{82}%
, 195312 (2010).

\bibitem{Savenko} I. G. Savenko \textit{et al}, Phys. Rev. B \textbf{83},
165316 (2011).

\bibitem{Ostatnicky} T. Ostatnicky, I.A. Shelykh and A.V. Kavokin, Phys.
Rev. B \textbf{81}, 125319 (2010).

\bibitem{ivchenkopikus} E.L. Ivchenko and G.E. Pikus,  \emph{Superlattices
and other heterostructures} (Springer-Verlag,  Berlin, 1997).

\bibitem{enaki} N. A. Enaki and O. B. Prepelitsa, Theoretical and
Mathematical Physics \textbf{88}, 967 (1991).

\bibitem{Ivchenko} E.L. Ivchenko, \emph{Optical Spectroscopy of
Semiconductor Nanostructures} (Alpha Science, Harrow UK, 2005).
\end{thebibliography}
\end{document}